\documentstyle[preprint,aps]{revtex}

\def\by#1#2{{\displaystyle {#1}\over \displaystyle {#2}}}
\def\d{{\rm d}}
\preprint {IMSc/99/05/22}
\begin{document}
\draft
\title{Neutrinos from Stellar Collapse: Effects of flavour mixing}

\author{Gautam Dutta, D. Indumathi, M. V. N. Murthy and G. Rajasekaran}

\address
{The Institute of Mathematical Sciences, Chennai 600 113, India.\\
}
\date{\today}
\maketitle
\begin{abstract}

We study the effect of non-vanishing masses and mixings among neutrino
flavours on the detection of neutrinos from stellar collapse by a water
Cerenkov detector. We consider a realistic frame-work in which there
are three neutrino flavours whose mass squared differences and mixings
are constrained by the present understanding of solar and atmospheric
neutrinos. We also include the effects of high dense matter within the
supernova core. We find that the number of events due to the dominant
process involving electron-antineutrinos may change dramatically for
some allowed mixing parameters.  Furthermore, contributions from
charged-current scattering off oxygen atoms in the detector can be
considerably enhanced due to flavour mixing; such events have a
distinct experimental signature since they are backward-peaked.

\end{abstract}


\narrowtext

\section{Introduction}

The sighting of a supernova, SN1987a in the large Magellanic cloud
(LMC) \cite{Announce}, led to great excitement, since, for the first
time, the neutrinos from stellar collapse were detected by earth--based
detectors \cite{KII,IMB}.  Unlike electromagnetic radiation, which
takes a long time to emerge from the collapsing core, neutrinos provide
direct information about core collapse.  The direct observation of
neutrinos from SN1987a by the Kamiokande (KII) \cite{KII} and the IMB
\cite{IMB} detectors forms the beginning of a new phase in neutrino
astrophysics with far--reaching implications for particle physics.
Since then, the KII detector has been upgraded with tremendous
improvement both in size and resolution, and many new detectors like
SNO and Borexino will begin taking data soon.

Immediately after SN1987a, several authors analysed
\cite{Dass,Arafune,Bahcall,Sato,Arnett,Kolb,Cow} the neutrino events
recorded by the KII and the IMB detectors. While the number of events
were not statistically significant enough to obtain quantitative
information on the neutrino spectrum, there was a qualitative agreement
between the predictions from the core collapse mechanism and the
observations. The present situation with improved neutrino detectors
affords quantitative analysis of the neutrino events if a supernova
collapse were to take place in the near future. While the observational
scenario is positive, there has also been much progress in
understanding the properties of neutrinos, namely their masses and
mixings, through the analyses of solar and atmospheric neutrino puzzles
\cite{Nurev}. Both the solar and atmospheric neutrino deficits that are
observed conclusively point to the requirement of (at least) three
neutrino generations and mixing among them. The masses and mixings are
then constrained by the observed deficit in the neutrino fluxes
\cite{Mohan,Fogli}.

In this paper, we analyse in detail the signatures of neutrinos from
stellar collapse. The analysis is confined to Type II supernovae (which
occur when the initial mass of the star is between 8--20 solar masses)
since the neutrino emission from these are significant enough to make
reasonable predictions. We consider the emission of all three types of
neutrino (and antineutrino) flavours. There exist many models of
stellar collapse.
The present understanding of neutrino emission involves dividing the
neutrino emission into two distinct phases---the neutronisation burst
and thermal neutrino emission. The number of neutrinos emitted during
the burst phase is only a few percent of the neutrinos emitted during
the thermal or cooling phase of the proto-neutron star. While only
$\nu_e$ is emitted during the earlier phase, neutrinos and
antineutrinos of all types are emitted during the final phase. Much of
the binding energy of the neutron star is radiated away as neutrinos
while a small fraction (few percents or less) is deposited in the shock
wave that blows away the mantle \cite{BL}. Detailed predictions for the
luminosity and average energy as a function of time are available
\cite{Totani} for neutrinos emitted during the burst and the cooling
phase. We use these predictions as an input in our analysis.

An important fact to note is that the neutrinos, which are produced in
the high dense region of the core, interact with matter before emerging
from the supernova. The presence of non-zero masses and mixing in
vacuum among various neutrino flavours results in strong matter
dependent effects, including conversion from one flavour to another.
Hence, the observed neutrino flux in the detectors may be dramatically
different for certain neutrino flavours, for certain values of mixing
parameters, due to neutrino oscillations. The effect of mixing on the
neutrino signal from supernovae has been analysed in detail before by
several authors \cite{KuoP,QF,Kar}. The effect of masses and mixing on
time-of-flight information has been discussed in \cite{BV,Kar2}.

Our analysis is based on the pioneering work of Kuo and Pantaleone
\cite{KuoP} where they include mixing among all three neutrino
flavours. However, unlike all the previous analyses, we take into
account the constraints on the neutrino mixing and masses imposed
\cite{Mohan,Fogli} by solutions consistent with the solar and
atmospheric neutrino puzzles. There are several possible solutions
here, including a purely vacuum solution for the solar neutrino
problem. We choose some typical (allowed) values for the mixing angles,
to illustrate the possible effects of mixing. This is important from
the point of view of integrating known constraints on the neutrino
masses and mixings in order to obtain a realistic picture of neutrino
emissions from supernovae.  We analyse in detail the dependence of the
recoil energy spectra on the mixing parameters, at water Cerenkov
detectors. In Sec. II we review the theoretical framework of our
calculation, including a reanalysis of the matter effects on the
neutrino spectrum in the hot dense core.  Sec. III highlights the
various inputs---neutrino fluxes and cross-sections in the
detector---that we have used in order to compute the event rate
expected at water Cerenkov neutrino detectors. The numerical
computation of the total number of events for the time integrated
neutrino spectrum, and the effects of neutrino oscillation, are
discussed in Sec. IV, where the results are discussed and summarised.
Some well-known results, adapted to the present situation, are
discussed in Appendices A and B for completeness.

\section{Three flavour oscillations in high dense matter} 

In this section, we discuss the mixing among three flavours of
neutrinos (or antineutrinos) and compute the electron (or antielectron)
neutrino survival probability, $P_{ee}$ (or $\overline{P}_{ee}$). We
will explicitly show that these are the only relevant probabilities.
While the theoretical details of this mixing are well-known, the
effects of super-dense matter, such as is found in the stellar cores,
are non-trivial. We shall also use this analysis to set our notation.

The three flavour eigenstates are related to the three mass eigenstates
in vacuum through a unitary transformation,
\begin{equation}
\left[ \begin{array}{c} \nu_e \\ \nu_{\mu} \\ \nu_{\tau}
\end{array} \right] = U^v 
\left[ \begin{array}{c} \nu_1 \\ \nu_2 \\ \nu_3 
\end{array} \right],
\end{equation}
where the superscript $v$ on the r.h.s. stands for vacuum.  The $3
\times 3$ unitary matrix, $U^v$, can be parametrised by three Euler
angles $(\omega, \phi, \psi)$ and a phase. The form of the unitary
matrix can therefore be written, in general, as
$$
U^{v} = U_{23}(\psi)\times U_{phase}\times U_{13}(\phi)\times
U_{12}(\omega)~,
$$ 
where $U_{ij}(\theta_{ij})$ is the mixing matrix between the $i^{th}$
and $j^{th}$ mass eigenstates with the mixing angle $\theta_{ij}$. It
has been shown that the expression for electron neutrino survival
probability, integrated over the time of emission and of absorption, is
independent of the phase and the third Euler angle $\psi$
\cite{KuoP87,AJMM88}. They can be set to zero without loss of
generality and we have the following form for $U^v$ :
\begin{equation}
U^v = \left( \begin{array}{ccc}
      c_{\phi} c_{\omega} & c_{\phi} s_{\omega} & s_{\phi} \\
      -s_{\omega} & c_{\omega} & 0 \\
      -s_{\phi} c_{\omega} & -s_{\phi} s_{\omega} & c_{\phi} \\
      \end{array} \right),
\end{equation}
where $s_{\phi} = \sin \phi$ and $c_{\phi} = \cos \phi$, etc. The
angles $\omega$ and $\phi$ can take values between $0$ and $\pi/2$.
Recently, the {\sc chooz} collaboration set a laboratory limit on
$\overline{\nu}_e$ oscillations \cite{Chooz}
that resulted in a strong limit, $\phi < 12^\circ$, on the (13) mixing
angle \cite{MRU}. However, a combination of solar and atmospheric
neutrino data allows for both large and small angle solutions for the
(12) mixing angle, $\omega$. The angle $\psi$ is large, typically of the
order of $\psi \sim 45^\circ$ (although this is not relevant here). These
constraints will be imposed later on in our numerical analysis.


The masses of the eigenstates in vacuum are taken to be $\mu_1$,
$\mu_2$ and $\mu_3$.
In the mass eigenbasis, the $({\rm mass})^2$ matrix is diagonal:
\begin{eqnarray}
M_0^2  & = & \left( \begin{array}{ccc} 
	             \mu_1^2 & 0 & 0 \\
                     0 & \mu_2^2 & 0 \\
	             0 & 0 & \mu_3^2  \\ 
		     \end{array} \right) \nonumber \\
 & = & \mu_1^2 I\!\!I + \left( \begin{array}{ccc}
			 0 & 0 & 0 \\
			 0 & \delta_{21} & 0 \\
			 0 & 0 & \delta_{31} \\
			 \end{array} \right),  \nonumber \\
 & = & \mu_1^2 I\!\!I + \Delta M_0^2~,  
\end{eqnarray}
where the mass squared differences are given by $\delta_{21} = \mu_2^2
- \mu_1^2$ and $\delta_{31} = \mu_3^2 - \mu_1^2$. Without loss of
generality, we can take $\delta_{21}$ and $\delta_{31}$ to be greater
than zero; this defines the standard hierarchy of masses. Neutrino
oscillation amplitudes are independent of the first term so we drop it
from further calculation. In the flavour basis, therefore, the relevant
part of the mass squared matrix has the form,
\begin{eqnarray}
\Delta M_v^2  & = & U^v \, \Delta M_0^2 \, {U^v}^{\dagger} \nonumber \\
       & = & \delta_{31} M_{31} + \delta_{21} M_{21}, 
\label{eq:mass}
\end{eqnarray}
where
\begin{eqnarray}
M_{31} & = & \left( \begin{array}{ccc} 
	s_{\phi}^2 & 0 & s_{\phi} c_{\phi}~; \\
	0 & 0 & 0 \\
       	s_{\phi} c_{\phi} & 0 & c_{\phi}^2 \\
         \end{array} \right)~; \nonumber  \\
M_{21} & = &  \left( \begin{array}{ccc}
      c_{\phi}^2 s_{\omega}^2 & c_{\phi} s_{\omega} c_{\omega} &
      -c_{\phi} s_{\phi} s_{\omega}^2 \\
      c_{\phi} s_{\omega} c_{\omega}  & c_{\omega}^2 & 
      -s_{\phi} s_{\omega} c_{\omega} \\
      -c_{\phi} s_{\phi} s_{\omega}^2 & -s_{\phi} s_{\omega} c_{\omega}
       & s_{\phi}^2 s_{\omega}^2 \\ \end{array} \right)~. 
\label{eq:Mij}
\end{eqnarray}
The relevant matter effects may  be included by a modified mass squared
matrix,
\begin{equation}
\Delta M_m^2 = \delta_{31} M_{31} + \delta_{21} M_{21} + A(r) M_A,
\label{eq:mmsq}
\end{equation}
where
\begin{equation}
M_A = \left( \begin{array}{ccc} 1 & 0 & 0 \\ 0 & 0 & 0 \\ 0 & 0 & 0 \\ 
       \end{array} \right)~,
\label{eq:A}
\end{equation}
and $A(r)$ is given by
\begin{equation}
A(r) = \sqrt{2}~ G_F ~N_e (r) \times 2 E~, \label{eq:defA}
\end{equation}
which is proportional to the electron number density, $N_e(r)$, in the
supernova core. Here $r$ is the radial distance from the centre of the
star. The detailed modifications due to matter effects are discussed
in Appendix A.

The maximum value of $A$ occurs at the core and is approximately
$2\times10^{7} E \ {\rm eV}^2$, where $E$ is the neutrino energy in
MeV.  The modification due to the matter dependence is similar to the
case of solar neutrinos, although, unlike in the case of solar
neutrinos, all flavours are produced in the supernova core.

It is clear that the mass squared matrix is no longer diagonal in the
presence of matter; we therefore diagonalise $\Delta M_m^2$ in order to
determine the matter corrected eigenstates. This is a difficult problem
in general for arbitrary values of $\delta_{31}$ and $\delta_{21}$.
These are however constrained by the limits on them given by the
simultaneous analysis of solar and atmospheric neutrino problems,
namely $10^{-3}\le \delta_{31} \le 10^{-2}$ eV${}^2$ and $ \delta_{21}
< 10^{-4}$ eV${}^2$, so that $\delta_{31} \sim \delta_{32}$; the value
of $A$ for energetic neutrinos (of a few MeV to tens of MeV) in the
core is therefore several orders of magnitude greater than these
mass-squared differences.  The eigenvalue problem may thus be solved
perturbatively, with the following hierarchy:  $A (\hbox{core}) \gg
\delta_{31} \gg \delta_{21}$. As a result, the electron neutrino
undergoes two well-separated resonances when the value of $A(r)$
approaches the two mass squared differences. Following Kuo and
Pantaleone \cite{KuoP}, and using the above mass hierarchy, the matter
mixing angle $\phi_m$ is given by (see Appendix A for more details),
\begin{equation}
\tan 2 \phi_m  =  \frac{\delta_{31} \sin 2 \phi}{
		  \delta_{31} \cos 2 \phi - A }~. 
		  \label{eq:tphim}
\end{equation}
At the point of production inside the core, $A (\hbox{core}) \gg
\delta_{31}$; thus, $\phi_m \to \pi/2$. This makes further calculations
extremely simple, since the electron neutrino is produced as a pure
$\vert \nu_3 \rangle $ mass eigenstate in the core of the supernova.
The survival probability of the electron neutrino is simply given by
the projection of the $\vert \nu_3 \rangle$ mass eigenstate on to the
$\vert \nu_e \rangle$ flavour state in the detector, after correcting
for the Landau-Zener jumps which may occur in the stellar matter during
propagation. The average survival probability of the electron neutrino
is therefore given by
\begin{eqnarray}
P_{ee} &=& \sum_{i,j=1}^3 \left|U^v_{ei}\right|^2
    \left|U^m_{ej}\right|^2 \left| \langle \nu^v_i \left| \right.
    \nu^m_j \rangle \right|^2~,\nonumber\\ 
& =&\sin^2\phi \, P_3 + \cos^2\phi \sin^2\omega \, P_2 + \cos^2\phi 
\cos^2\omega \, P_1~. 
\end{eqnarray}
Here $\phi$ and $\omega$ are the vacuum mixing angles defined earlier and 
$P_i$ denote the Landau-Zener jump probabilities among the mass 
eigenstates,
\begin{eqnarray}
P_1 &=& P_h P_l~, \\
P_2 &=& P_h (1  - P_l)~, \\
P_3 &=& (1  - P_h)~,
\end{eqnarray}
where $P_h$ and $P_l$ denote the jump probabilities at the higher and at 
the lower resonances. In Appendix B we show that for the parameters 
values relevant in the case of neutrinos produced in the supernova core,
$P_h$ is actually very close to zero. Therefore to a good approximation, 
we may write,
\begin{equation}
P_{ee}=\sin^2\phi~. 
\end{equation}
This result implies that the propagation of neutrinos is adiabatic in
the high dense core. We will support this conclusion in Appendix B.
The information given above is not enough to obtain completely the
survival and oscillation probabilities of the individual flavours.
However, since the detectors we are interested in do not separately
detect $\nu_\mu$ and $\nu_\tau$, this is sufficient for our analysis.
We shall therefore use this form for the survival probability for the
numerical results calculated in the next section.

We now consider the case of $\overline{\nu}_e$ propagation in  high
dense matter. The only change in this case is that the matter dependent
term in the relevant part of the mass squared matrix has the opposite
sign (to that in Eq.~(\ref{eq:defA})), that is,
\begin{equation}
A(r) = -\sqrt{2}~ G_F ~N_e (r) \times 2 E~.
\end{equation}  
The analysis goes through as in the case of $\nu_e$ propagation through
matter and the mixing angle, $\phi$, for antineutrinos in matter is
given by,
\begin{equation}
\tan 2 \phi_m  =  \frac{\delta_{31} \sin 2 \phi}{
		  \delta_{31} \cos 2 \phi + A }. 
		  \label{eq:tphiam}
\end{equation}
On using the fact that $A ({\hbox{core}}) \gg \delta_{31}$, we obtain
$\phi_m \to 0$ in contrast to the solution $\phi_m \to \pi/2$ for
electron neutrinos.  Thus $\overline\nu_{e}$ is produced in the mass
eigenstate, $\vert \overline\nu_1 \rangle $, in the core of the
supernova.  There are no Landau-Zener jumps to consider in this case
since the resonance conditions are never satisfied unless the mass
hierarchy is altered.  The propagation is therefore adiabatic and the
survival probability is obtained by simply projecting the $\vert
\overline\nu_1\rangle $ eigenstate on to the flavour eigenstate in
vacuum (at the detector). The antineutrino survival probability is
therefore given by,
\begin{equation}
\overline P_{ee} =\cos^2\phi \cos^2\omega~, 
\end{equation}
where $\phi$ and $\omega$ are as usual the vacuum mixing angles defined 
earlier.

\section{Neutrino fluxes and cross-sections}

We need the inputs of neutrino flux emission at the supernova, and
neutrino cross-section at the detector, in order to obtain the event
rates. We begin with a discussion of the neutrino fluxes.

\subsection{Neutrino fluxes}

Following Kuo and Pantaleone \cite{KuoP}, we denote the flux of various
flavours of neutrinos and antineutrinos produced in the core of the
supernova by $F_i^0$, where $i$ denotes all the flavours. In particular
we use the generic label $F_x^0$ for flavours other than $\nu_e$ and
$\overline\nu_e$ since
\begin{equation}
F_x^0 = F_{\nu_{\mu}}^0
= F_{\overline\nu_{\mu}}^0
= F_{\nu_{\tau}}^0
= F_{\overline\nu_{\tau}}^0~.
\end{equation}
All these flavours are produced via the neutral-current (NC) pair
production processes and therefore have the same flux for all practical
purposes. However, the $\nu_e$ and $\overline\nu_e$ fluxes are
different from each other and the rest since they are produced not only
by pair production but also derive contribution from charged-current
(CC) processes.

In the presence of matter, the flux emerging from the core undergoes
changes due to oscillations as was discussed in the previous section.
The flux reaching the detector from a supernova at a distance $d$ from
earth is reduced by an overall geometric factor of $1/(4\pi d^2)$.
Apart from this, there is a further modification of the observed flux
due to oscillations. The flux on earth, in the various flavours, is
given in terms of the flux of neutrinos produced in the core of the
supernova by,
\begin{eqnarray}
F_{\nu_e} &=& P_{ee} F_{\nu_e}^0 +P_{e\mu}F_{\nu_{\mu}}^0 + 
P_{e\tau}F_{\nu_{\tau}}^0, \nonumber \\ 
&=& F_{\nu_e}^0 -(1-P_{ee})(F_{\nu_e}^0-F_x^0)~, 
\end{eqnarray}
where we have made use of the constraint $\sum_{j} P_{ij} =1$ and
$P_{ex}$ denotes the probability of a flavour $\mu$ or $\tau$ neutrino
emerging as an electron neutrino. Since $\nu_{\mu}$-- and
$\nu_{\tau}$--induced events cannot be separated in water Cerenkov
detectors, their combined flux on earth may be written as
\begin{eqnarray}
2 F_x &=& F_{\nu_{\mu}} + F_{\nu_{\tau}}~, \nonumber \\ 
&=& 2F_x^0 +(1-P_{ee})(F_{\nu_e}^0-F_x^0)~. 
\end{eqnarray}
Note that flavour mixing does not affect the total flux.

Similar expressions hold for antineutrino flavours with appropriate 
changes, that is,
\begin{equation}
F_{\overline{\nu}_e} = F_{\overline\nu_e}^0 -(1-\overline{P}_{ee})
(F_{\overline\nu_e}^0-F_x^0)~;
\end{equation}
and
\begin{equation}
2 F_{\overline{x}} = 2F_x^0 +(1-\overline{P}_{ee})
      (F_{\overline\nu_e}^0 -F_x^0)~.
\end{equation}
Since $P_{ee} \neq \overline{P}_{ee}$, in general, the mixing breaks
the equality of the $\nu_x$ and $\overline{\nu}_x$ fluxes.

We use the luminosity and average energy distributions (as functions of
time) as given in Totani et al. \cite{Totani}, based on the numerical
modelling of Mayle, Wilson and Schramm \cite{BL}. The neutrino number
flux is described, in a given time interval, $\Delta t$, as a thermal
Fermi Dirac distribution,
\begin{equation}
\by{\d F_i^0(j)}{\d E} = N_0 \by{{\cal L}_i}{T_j^4} \by{E^2}{(\exp(E/T_j)
                                 +1)} ~,
\end{equation}
for neutrinos of flavour $j$ and energy $E$ at a time $t$ after the
core bounce. Here $i$ refers to the time-bin, $t = t_0 + i \Delta t $.
Hereafter, we set the time of bounce, $t_0 = 0$. The overall
normalisation, $N_0$, is fixed by requiring that the total energy
emitted per unit time equals the luminosity, ${\cal L}_i$, in that time
interval.

The thermal distribution that we have used (where the chemical
potential has been set to zero) results in a flux that shows a slower
fall with energy, $E$, than the results of the corresponding numerical
model used in Ref.~\cite{Totani}; however, the effect of this on the
event rates is small, of the order of a few percent.  Typically, this
distribution corresponds to an average neutrino energy or temperature
($\langle E_j \rangle \simeq 3.15 T_j$) of $\langle E (\nu_e)\rangle
\sim 12$ MeV, $\langle E (\overline{\nu}_e) \rangle \sim 16$ MeV, and
$\langle E (\nu_x) \rangle \sim 24$ MeV. Beyond about 1 second after
the bounce, the average energies remain constant over the emission
times of the supernova. However, the luminosities decrease, with very
little emission beyond 10 seconds. Hence, in order to compute the event
rates, we consider neutrino emission up to 10 seconds after bounce. The
total emitted energy in all flavours of neutrinos up to this time is
about $2.7 \times 10^{53}$ ergs, which is more or less equally distributed
in all flavours. The number of neutrinos emitted in each flavour,
however, is not the same since their average energies are different.

\subsection{Interaction at the detector}

The basic quantity we are interested in is the distribution of events
in the detector as a function of the energy of the detected particle.
In the case of a water Cerenkov detector, this corresponds to the
detection of a charged lepton in the final state. Here we are concerned
with detection of electrons (or positrons) with energy, $E_e$.  The
various processes of interest therefore are the interactions of the
neutrinos
\begin{enumerate}
\item with electrons in water as targets:
\begin{equation}
\nu_\ell (\overline{\nu}_\ell) + e^- \to
\nu_\ell (\overline{\nu}_\ell) + e^-~, \ell = e, \mu, \tau~; \\
\end{equation}

\item with free protons in water as targets:
\begin{equation}
\overline{\nu}_e + p \to e^+ + n~; \\
\end{equation}

\item with oxygen nuclei in water as targets:
\begin{eqnarray} \nonumber
\nu_e + {}^{16} O \to e^- + {}^{16} F~, \\
\overline{\nu}_e + {}^{16} O \to e^+ + {}^{16} N~.
\end{eqnarray}
\end{enumerate}
The cross-sections, $\d\sigma/\d E_e$,  for all these processes, except
the ones on oxygen, are well known \cite{tHooft,Bahcallb}. The oxygen
cross-sections have been taken from Fig.~1 of Haxton \cite{Haxton}. As
the interactions on protons and oxygen nuclei are purely CC
interactions, they involve only $\nu_e$ and $\overline{\nu}_e$.
Reaction (1) involves both CC and NC interactions for $\nu_e$ and
$\overline{\nu}_e$ and only NC interactions for all other flavours.

The $\overline{\nu}_e p$ cross-section is the largest, being
proportional to the square of the antineutrino energy. In terms of
total number of events, therefore, water Cerenkov detectors are mostly
dominated by $\overline{\nu}_e$ events. However, the different
interactions in the detector have distinct angular signatures; this may
be used to distinguish them. The elastic electron cross-sections are
forward peaked, especially for neutrinos with energies
\raisebox{-0.1cm}{$\stackrel{\displaystyle >}{\sim}$} 10 MeV
\cite{Bahcallb,GM}, while the proton cross-section is isotropic in the
lab frame. Finally, the CC $\nu_e$ ($\overline{\nu}_e$) cross-section
on oxygen, although having a rather large threshold of 15.4 MeV (11.4
MeV) \cite{Haxton}, increases rapidly with incoming neutrino energy and
is somewhat backward peaked. The higher the temperature at which the
neutrino is emitted, the larger is this backward peak; hence it may be
possible to distinguish this contribution from the rest by angular
resolution as well, especially if there is substantial mixing between
$\nu_e$ (or $\overline{\nu}_e$) and $\nu_x$ since the latter have a
considerably hotter spectrum.


\subsection{Event rates}

The time integrated event rate, from neutrinos of flavour
$j$ and energy $E$, as a function of the recoil electron (or positron)
energy is given by,
\begin{equation}
\by{\d N^t(j)}{\d E_e} = \by{N_t}{4\pi d^2} \sum_i \int \d E 
		       \by{\d F_i (j)}{\d E} \by{\d \sigma_p}{\d E_e}~,
\label{eq:rate}
\end{equation}
where the flux distribution, $\d F (j)/\d E$ includes the effects of
mixing in the hot dense core and the index $t$ refers to any of the
various processes through which the neutrino $j$ can interact with the
detector. Here $N_t$ refers to the number of scattering targets (of
either $e$, $p$ or ${}^{16}O$) that are available in the detector. The
total number of events from a given flavour of neutrino in a given bin,
$k$, of electron energy (which we choose to be of width 1 MeV) then is
the sum over all possible processes integrated over the bin width of
the event rate:
\begin{equation}
N (j, k) =  \sum_t \int_k^{k+1} \d E_e \by{\d N^t(j)}{\d E_e}~.
\end{equation}
In the next section, we shall use this formula to compute the time
integrated event rates for neutrino scattering with and without mixing,
in water Cerenkov detectors, as a function of the detected electron (or
positron) energy, in order to examine the effects of neutrino
oscillations on supernova neutrino fluxes.

\section{Results and Discussion}
We compute the time integrated event rate at a prototype 1 Kton water
Cerenkov detector from neutrinos emitted by a supernova exploding 10
KPc away.  Results for any other supernova explosion may be obtained by
scaling the event rate by the appropriate distance to the supernova and
the size of the detector, as shown in Eq.~(\ref{eq:rate}).  We assume
the efficiency and resolution of such a detector to be perfect.
Including these effects does not change the results by more than a few
percents, as we will see. In fact, the maximum variation is at low
energies, close to the threshold, where the low detector efficiency
leads to lower detection rates.

The following constraints derived from solar and atmospheric neutrino
observations are imposed. We begin with the constraints in the neutrino
sector. Here, the angle $\omega$ does not play a role.  As stated
earlier, the (13) mixing angle is severely restricted: $\phi <
12^\circ$. The solar and atmospheric neutrino problems allow for a
wider choice in $\phi$. This restriction on $\phi$ comes mainly from
the {\sc chooz} experiment \cite{Chooz}.  Since the survival
probability, $P_{ee}$ depends only on this angle ($P_{ee} = \sin^2
\phi$); this implies that $P_{ee} < 0.05$ and is thus very small. The
observed dynamics of electron--type neutrinos, therefore, is completely
driven by mu-- and tau--type neutrinos produced in the supernova:
\begin{eqnarray}
F_{\nu_e} & \simeq & F_x^0~; \nonumber \\ 
2 F_x & \simeq & (F_x^0 + F_{\nu_e}^0)~.
\end{eqnarray}
Hence, due to mixing, the original electron neutrino flux is virtually
replaced by the $\mu$ or $\tau$ neutrino flux. The cross-sections at
the detector increase with energy. Since the average energy of $\nu_x$
is of the order of $\sim 24$ MeV while that of $\nu_e$ is $\sim 11$
MeV, the effect of mixing and matter in the dense core is to
dramatically increase the number of events due to $e$--type neutrinos
while reducing the corresponding $\nu_x$ contribution.

We now discuss the antineutrino sector. While the same limits apply on
$\phi$, we now have to consider the limits on the $(12)$ mixing angle
$\omega$ as well. The constraints on $\omega$ mainly emerge from the
solar neutrino problem. (For a recent review, see Ref.~\cite{Bahcallr}).
The best global {\sc msw} fit gives $\delta_{12} \sim 10^{-6}$ eV${}^2$
and $\sin^2 2\omega = 5.5 \times 10^{-3}$. There is also a large angle
solution with {\sc msw} fit. For vacuum oscillations, the fit gives
$\delta_{12} \sim 10^{-10}$ eV${}^2$ and $\sin^2 2\omega = 0.75$. In
the present analysis, $\delta_{12} < 10^{-4}$ eV${}^2$, which is
consistent with both {\sc msw} and vacuum solutions. For $\omega$,
therefore, we choose two possible values, viz., $\omega$ small and
$\omega$ large. These two typical choices cover the extreme ends of the
possible effects of mixing in supernova neutrinos.

If $\omega$ is small (corresponding to the {\sc msw} solution to the
solar neutrino problem), then the antineutrino survival probability
becomes, $\overline{P}_{ee} = \cos^2\phi \cos^2\omega \to \cos^2 \phi
\to 1$, since $\phi$ is small.  This, in effect, is similar to the
no-mixing solution. The large angle solution allows for a near-maximal
mixing of $\omega \sim 45^\circ$; in this case, the survival
probability becomes, $\overline{P}_{ee} = \cos^2\phi \cos^2\omega \to
1/2$ and this corresponds to maximal mixing in the antineutrino sector.
Therefore we have,
\begin{eqnarray}
F_{\overline{\nu}_e} & \simeq & \by{1}{2} \left(F_{\overline{\nu}_e}^0 +
F_x^0 \right)~, \nonumber \\ 
2 F_{\overline{x}} & \simeq & \by{1}{2} \left
(3 F_x^0 + F_{\overline{\nu}_e}^0 \right)~.
\end{eqnarray}
In any case, we have the result that $\overline{P}_{ee}$
\raisebox{-0.1cm}{$\stackrel{\displaystyle >}{\sim}$} 0.5 for any
choice of $\omega$ when $\phi$ is small.  Hence, typically, the
antineutrino fluxes that reach the earth are combinations of
$\overline{\nu}_e$ and $\overline{\nu}_x$ fluxes. Again, since the
average energies of $\overline{\nu}_e$ and $\overline{\nu}_x$ are 15
and 24 MeV respectively, this results in an enhanced $\overline{\nu}_e$
event rate and a reduced $\overline{\nu}_x$ rate at the detector. It is
important to note that these flux mixings are energy independent. For
example, the energy spectrum of a given neutrino flavour produced in
the supernova is not altered during propagation; however, its flavour
content at the detector will depend on the extent of mixing. We shall
now probe the quantitative effects of these mixings on the observed
event rates.

We first consider the case where there is no neutrino mixing.  The
largest contribution comes from the $\overline{\nu}_e\,p$ interaction,
which has a cross-section proportional to the square of the
antineutrino energy. This is shown in the left--hand part of Fig.~1,
where the number of events, $N(k)$, in the k$^{th}$ bin is plotted
against the central values of the recoil electron energy in that bin.
In comparison, the $\overline{\nu}_e\,e$ contribution is negligibly
small.  However, this is not the case with the $\overline{\nu}_e\,O$
contribution, which though small, may be measurable at, say, the large
(32 KTon) SuperK detector.  It can be seen, though, that the total rate
is saturated by the proton interaction.  Neutrino mixing causes an
increase in the high energy event rates, as can be seen from the right
hand side of Fig.~1. Here, we have used typical values for $\omega$ ($=
45^\circ$) and $\phi$ ($ = 10^\circ$); this choice of $\omega$
maximises mixing effects. The other (small-angle) solution for $\omega$
is similar to the no-oscillation scenario shown on the left hand side
of the figure. Note that our three-flavour analysis precludes the
choice $\omega, \phi = 0$. For example, the result that $\nu_e$ starts
out as a pure $\nu_3$ mass eigenstate in the stellar core will not hold
if $\phi = 0$.

The corresponding results for $\nu_e$ events are shown in Fig.~2. While
the no--mixing contributions are negligible, it is seen that there is a
more than 10--fold increase in the event rate due to scattering off
oxygen.  The low no--mixing rate was because the average $\nu_e$ energy
is less than the threshold energy required for this reaction to
proceed. Mixing opens up this channel since there are now many more
$\nu_e$, originating as $\nu_x$ in the star, which are more energetic.
Since the backward peak in the $\nu_e\,O$ cross-section is more
pronounced for flux distributions at higher temperatures \cite{Haxton},
it may be possible to separate these events from the bulk of the
anti-electron neutrino events at the detector.

Finally, we see from Fig.~3 that there is a low-energy enhancement of
the $\nu_x$ and $\overline{\nu}_x$ rates upon mixing. Their
contribution, however, is still small, of the order of the $\nu_e\,e$
elastic scattering events. For comparison, all the contributions, with
and without mixing, are shown in Fig.~4. It is clear that the proton
absorption events are the largest, independent of mixing. However, it
is the $\nu_e\,O$ events which are most sensitive to the amount of
mixing, and are likely to be most important in furthering our
understanding of neutrino oscillations in vacuum and matter.

We recall that the proton absorption events are isotropic, the
scattering off oxygen is backward enhanced, while the elastic
scattering on electrons is mostly forward peaked. In fact, even the
elastic $\nu_e$ and $\nu_x$ (including antineutrinos) events may be
separated based on angular resolution \cite{GM}. Hence, it is likely
that a nearby supernova explosion (at a distance of about 10 KPc, say)
can yield information {\it independently} on the various neutrino
flavours, $\nu_e$, $\overline{\nu}_e$ and $\nu_x$. We have therefore
shown the total contribution from each of these flavours, with and
without neutrino mixing, in Fig.~5. For a 32 KTon detector such as
SuperK, this translates to a total event rate of 12,235 events with
mixing as opposed to 9,871 events without mixing, a 25\% increase, with
individual channels contributing as shown in Table~1. Note that the
thermal flux distribution, while agreeing with the numerical model of
Ref.~\cite{Totani} at lower energies, overestimates the flux at larger
energies. Hence, the number of events at high energies may be
overestimated in this model. However, we emphasise that the
{\it relative} increase, with and without oscillation, remains the same.

We now briefly discuss these results in relation to the supernova
SN1987A. Recall that the supernova, which was 55 KPc away, was detected
by KII, which was a 2.14 Kton (fiducial volume) water Cerenkov
detector; the corresponding results for this can therefore be obtained
from our analysis by multiplying the results by a factor of
$2.14/30.25$. However, since KII mostly detected low energy events, we
have now included the detector efficiency \cite{QF} in our analysis.
The event rate is determined entirely by the $\overline{\nu}_e\, p$
events.  Since KII measured the time dependence of the spectrum, we
have shown our results for the event rate as a function of time in
Fig.~6. The figure on the left hand side of Fig.~6 shows the event
rates for energies from $8 < E_e \hbox{ (MeV)} < 30$ MeV, which is the
energy range in which KII made observations; the dotted curve indicates
the contribution in the absence of mixing. It is seen that mixing
marginally decreases the event rate while it almost doubles the rate
for events with energies $E_e > 30$ MeV as can be seen from the figure
on the right. This separation has been done since the thermal
Fermi-Dirac distribution that we use overestimates the flux at larger
(neutrino) energies due to a very large high-energy tail compared to the
corresponding numerical model \cite{Totani}, while agreeing quite well
at lower energies, $E_{\bar{\nu}_e} \sim E_e < 30$ MeV. This is
particularly true for the event rate at early times; hence our high
energy predictions may be overestimated by a factor of 4 or more. Note,
however, that even if the absolute spectrum is overestimated by the
model we have used, the results we had shown earlier contrast the
relative differences with and without mixing and still hold. Finally,
the model also predicts that high energy events are most likely to
occur at early times in the supernova explosion. This is not
inconsistent with the observations of KII \cite{KII}.

To summarise, a great deal of the physics of neutrino mixing and the
effects of dense matter in neutrino propagation may be tested from
neutrinos emitted during supernova explosions. Specifically, from the
results summarised in Table~1, we may conclude,
\begin{enumerate}
\item The observed $\overline{\nu}_e\,p$ events are the largest in
number as well as least sensitive to the mixing parameters. Hence they
provide a direct test of the supernova models. Since the angular
distribution of these events is isotropic, they may be used to set the
overall normalisation.  
\item All the interactions involving electrons as targets are peaked in
the forward direction (in fact, for $E_\nu > 8$ MeV, more than 90\% of
them lie in a $10^\circ$ cone with respect to the supernova direction).
In the absence of any mixing, there will also be a few events in the
backward direction due to CC scattering on oxygen targets. As indicated
in Table~1, the forward-backward asymmetry in the event distribution
will be clearly marked.
\item The main effect of mixing is then to produce a dramatic increase
in the events involving oxygen targets. As remarked earlier, this will
show up as a marked increase in the number of events in the backward
direction with respect to the forward peaked events. The actual
increase, however, will depend sensitively on a combination of both the
mixing parameters as well as the supernova model.
\end{enumerate}
We have limited our analysis in this paper to a model with three active
neutrino generations and possible mixings among them. This allows us to
incorporate constraints arising from solar and atmospheric neutrino
problems. This however leaves out a new set of constraints which may
emerge from the results of the {\sc lsnd} experiments \cite{LSND}. The
{\sc lsnd} results cannot be accommodated within the three generation
formalism, with the parameter ranges used in our analysis. One may
therefore require a sterile neutrino with a new mass scale, leading to
yet another mass squared difference, $\delta \sim 0.3$--1 eV${}^2$
\cite{Rev}.  In the context of supernova neutrinos, this opens up yet
another channel for $\nu_{\mu,\tau}$ to oscillate and may therefore
reduce the dramatic enhancement one sees in the $\nu_e$ events.

We thank Kamales Kar for many useful discussions and a critical reading
of the paper.

\newpage
\section*{Appendix A}
The time evolution of the neutrino mass eigenstates in vacuum is given
by,
$$
\nu_i(t) = \exp[ - i E_i t] \nu_i(0)~; i = 1, 2, 3.
$$
Assuming the neutrino masses to be small, in the extreme relativistic
limit, we have,
$$
E_i \simeq p + \by{\mu_i^2}{2p}~,
$$
where $\mu_i, (i = 1, 2, 3)$, denotes the neutrino masses. In the
presence of matter, neutrinos interact with electrons, protons and
neutrons in matter. While $\nu_e$ ($\overline{\nu}_e$) interact both via
CC and NC interactions, $\nu_x$ ($\overline{\nu}_x$) scatter via NC
interactions alone. Note that interaction with matter is diagonal in the
flavour basis but not in the mass basis. As a result, the dispersion
relation in matter is given by,
$$
E_i \simeq p + \by{m_i^2}{2p}~,
$$
where $m_i$ are now eigenvalues of the (mass)${}^2$ matrix given by
$$
M_m^2 = M_v^2 + M_{\rm int}^2~.
$$
Here the mass-squared matrix in vacuum, $M_v^2$, is defined in
Eqs.~(3) and (\ref{eq:mass}) and $M_{\rm int}$ is given by 
$$
M_{\rm int}^2 = \pm \sqrt{2} G_F p \left[ - \left\{(1-4\sin^2\theta_W)
    (N_e -N_p) + N_n\right\} I\!\!I + 2 N_e M_A \right]~.
$$
Here $N_e$, $N_p$ and $N_n$ denote the number densities of electrons,
protons and neutrons in matter, $M_A$ is the matrix defined in
Eq.~(\ref{eq:A})
of Sec.~2, and $\theta_W$ is the Weinberg angle. Note that
$(1-4\sin^2\theta_W)$ is close to zero and that the matrix,
$M_{\rm int}^2$, is expressed in the flavour basis. The upper sign
corresponds to neutrinos ($\nu_e$, $\nu_\mu$ and $\nu_\tau$) while the
lower sign corresponds to antineutrinos ($\overline{\nu}_e$,
$\overline{\nu}_\mu$, $\overline{\nu}_\tau$).

We will now compute the eigenvalues, $m_i^2$, in matter, ignoring
the terms proportional to the unit matrix, $I\!\!I$, since we will only be
interested in differences between the eigenvalues. Consider the
eigenvalues of the matrix defined by
$$
\Delta M_m^2 \equiv A(r) M_A + \delta_{31} M_{31} + \delta_{21} M_{21}~,
$$
where $\delta_{ij} = \mu_i^2 - \mu_j^2$ and $M_{31}$, $M_{21}$ are
matrices defined in Eq.~(\ref{eq:Mij}) and $A(r)$, the matter dependent
(and hence distance dependent) term, $A(r) = \pm 2\sqrt{2} G_F N_e E$
for $\nu_e$ and $\overline{\nu}_e$ respectively, varies linearly with
the matter density. Since neutrinos are produced in the high dense
region of the supernova, $\vert A (\rm core) \vert \simeq 2 \times
10^7$ E eV${}^2$ where $E$ is the neutrino energy in MeV. Thus $\vert A
(\rm core) \vert \gg \delta_{31} \gg \delta_{21}$.

We now compute the eigenvalues perturbatively. Since $\delta_{31}$ and
$\delta_{21}$ are different from each other, the resonances, if they
occur, are well-separated. We therefore diagonalise the first two terms
in $\Delta M_m^2$ and treat the third term as a perturbation. The
eigenvalues are then given by,
\begin{eqnarray}
\delta m_1^2 & = & \by{A+\delta_{31}}{2} + \by{1}{2} \left[ (A -
\delta_{31} \cos 2\phi)^2 + (\delta_{31} \sin 2\phi)^2 \right]^{1/2} +
\delta_{21} \cos^2(\phi - \phi_m) \sin^2 \omega ~; \\
\delta m_2^2 & = &  \delta_{21} \cos^2 \omega~; \\
\delta m_3^2 & = & \by{A+\delta_{31}}{2} - \by{1}{2} \left[ (A -
\delta_{31} \cos 2\phi)^2 + (\delta_{31} \sin 2\phi)^2 \right]^{1/2} +
\delta_{21} \sin^2(\phi - \phi_m) \sin^2 \omega ~;
\end{eqnarray}
where the matter mixing angles are given by
$$
\tan 2 \phi_m = \by{\delta_{31}\sin2 \phi}{\delta_{31} \cos 2 \phi - A}~,
$$
and 
$$
\tan \omega_m = {\cal O} \left(\by{\delta_{21}}{A}
\right)~.
$$
To the leading order, the mixing matrix in matter, $U_m (\phi_m, \omega_m)$,
is given by,
$$
U_m = \left[ \begin{array}{ccc}
c_{\phi_m} & \Lambda s_{\phi_m} & s_{\phi_m} \\
0 & 1 & -\Lambda \\
-s_{\phi_m} & \Lambda c_{\phi_m} & c_{\phi_m} \\ \end{array} \right]~,
$$
which is unitary, up to ${\cal O} ((\delta_{21}/\delta_{31})^2)$~. Here
$c$ and $s$ stand for $\cos$ and $\sin$ respectively, for example,
$c_{\phi_m}$ denotes $\cos \phi_m$; $\Lambda = (\delta_{21}/\delta_{31})
s_{(\phi-\phi_m)} s_\omega c_\omega$. Terms of order ${\cal{O}}
(\delta_{21}/A)$ are neglected in $U_m$. 

Uptil now, the only approximation that has been used is the hierarchy,
$A \gg \delta_{31} \gg \delta_{21}$. Using this hierarchy and the value
of $A (\rm core)$ given earlier, we find,
$$
\phi_m \to \left\{ \begin{array}{cl} 
\pi/2 & \hbox{for neutrinos}~; \\
0 & \hbox{for antineutrinos.} \\
\end{array} \right. 
$$
This result, when combined with the definition of the matter mixing matrix,
$U_m$, leads to the fact that $\nu_e$'s are produced almost entirely in
the $\vert \nu_3 \rangle$ mass eigenstate whereas $\overline{\nu}_e$
are produced almost entirely in the mass eigenstate, $\vert
\overline{\nu}_1 \rangle$, in the dense stellar core.

If the propagation is adiabatic, this also implies that the averaged
survival probability of electron neutrinos is given by
$$
P_{ee} = \left\vert \langle \nu_e(t) \vert \nu_3(0) \rangle
\right\vert^2 = s_\phi^2~,
$$
and that of the anti-electron neutrino is given by
$$
\overline{P}_{ee} = \left\vert \langle \overline{\nu}_e(t) \vert
\overline{\nu}_1(0) \rangle
\right\vert^2 = c_\phi^2 c_\omega^2~,
$$
where $t$ denotes the time of detection of the neutrino on earth.

In Appendix B we show that the propagation is indeed adiabatic and the
expressions given above provide a reasonably accurate description of the
matter effects in the stellar interior. 

A few remarks about the eigenvalues, $\delta m_i^2$, are in order. Note
that the eigenvalues themselves are always positive definite for
electron type neutrinos, whereas for muon type neutrinos or
electron-antineutrinos, this is not always the case since $A$ is large
and negative. The complete dispersion relation for the energy
eigenvalues is given by
$$
E_i \simeq p + \by{1}{2p} \left[ \mu_i^2 \mp \sqrt{2} G_F p \left\{ (1 -
4\sin^2\theta_W)(N_e - N_p) + N_n \right\} + \delta m_i^2\right]~.
$$
The effect of the CC interactions with matter gives rise to the $\delta
m_i^2$ term. The NC term, common to all flavours, is now included here.
The second term in the expression for $E_i$ is typically of the order
of tens of eV in the stellar core. Therefore, for energies $E \sim p
\sim$ few MeV, the second term is small and may be neglected except
when computing matter mixing angles.  However, for neutrinos having
energies of the order of tens of eV (but still with $p \gg \mu_i$), the
two terms compete. Since the sign of the second term changes depending
on whether the particle scattering is a neutrino or an antineutrino,
one may expect interesting phenomena when $E_i$ becomes negative. This
may lead to the trapping of low-energy neutrinos. This phenomenon is
unique to neutrinos produced in supernova explosions.  While it is of
little relevance to the detection of neutrinos on earth, it may have
interesting astrophysical consequences.  The dynamics of such neutrinos
is under further investigation.

\section*{Appendix B}

The electron neutrino $\nu_e$ is produced in the core of the supernova in
the mass eigenstate $\vert \nu_3 \rangle$ with a negligible admixture of
the other two states. As the produced $\vert \nu_3 \rangle$ propagates
outwards, it passes through variable density matter (since the density
inside the stellar core decreases monotonically outwards). Such a
propagation may in
general induce the presence of other mass eigenstates (since they are no
longer eigenstates of the Hamiltonian) by the time the neutrino exits
the star and reaches the detector. The Landau-Zener ``jump probability''
or level transition probability \cite{LZ} is maximal at the resonances
and is given by
$$
P_{LZ} = \exp \left[ - \by{\pi}{2} \gamma F \right]~,
$$
where $F$ is a factor which depends on the density profile and $\gamma$
is the non-adiabaticity parameter. 

Since $\nu_e$ is produced in the mass eigenstate $\vert \nu_3 \rangle$,
we first consider the cross-over between $\vert \nu_3 \rangle $ and
$\vert \nu_2 \rangle $. Then, 
$$
\gamma = \by{\delta \sin^2 2 \phi}{2E \cos 2 \phi \left\vert
\frac{1}{N_e} \frac{\d N_e}{\d x} \right\vert_0}~,
$$
where $\phi$ is the relevant mixing angle for the upper resonance and
$\delta \equiv (\delta_{31} + \delta_{32} +
\delta_{12} \cos2\omega)/2 \sim \delta_{31}$ independent of $\cos2\omega$
\cite{KuoP}, because of the assumption, $\delta_{12} \ll \delta_{31} \sim
\delta_{32}$. The suffix `0' indicates that the derivative in the
density, $N_e$, is to be evaluated at resonance, when the eigenstates
$\vert \nu_2 \rangle$ and $\vert \nu_3 \rangle$ are the closest.

The density profile in the core may be assumed to be of the form \cite{KuoP},
$$
\rho (r) \sim \by{C}{r^3}~,
$$
where $1 < C/10^{31} \hbox{gms} < 15$. (This assumption is not crucial
but is sufficient for our analysis). As a result, the non-adiabaticity
parameter evaluates to
$$
\gamma = \by{R_0 \delta_{31}}{6E} \left( \by{\sin^22\phi}{\cos2\phi}
\right) ~,
$$
where $R_0$ is the radius at which the higher resonance occurs,
i.e., when $ A(r) \simeq \delta_{31} \cos2\phi$. Using the explicit
expression for $A(r)$, the resonant density is given by
$$
\rho_0 \hbox{(gm/cc)} = \by{6.6 \times 10^5}{Y_e} \left(\by{\delta_{31}}
{e}\right)\cos2\phi~,
$$
where $\delta_{31}$ is in eV${}^2$ and $e = E/(10$ MeV). Here
$Y_e = Z/A \sim 0.5$ is the electron fraction in the matter. 
For $\delta_{31} \simeq 10^{-3} \hbox{ eV}^2$, as preferred by solar
and atmospheric neutrino data, and a typical detected neutrino energy of
$10$ MeV, the resonant density is $\rho_0 = 1320 (\cos2\phi)$ gm/cc. This
determines the resonant radius, $R_0$ for a given value of
$\phi$. If the angle $\phi$ is indeed small, as indicated by {\sc chooz}
\cite{Chooz}, this implies 
$$
R_0 = \left( \by{eC}{1320 \times 10^{15}} \right)^{1/3}~ \hbox{km}~,
$$
which evaluates to 20--50 thousand kilometres for $E = 10$ MeV ($e =
1$).

The non-adiabaticity parameter is then given by (for $R_0$ in km),
$$
\gamma \simeq \by{5076 R_0}{e}
\left( \by{\sin^22\phi}{\cos2\phi} \right)~.
$$
Since $R_0$ is large, it is clear that $\gamma$ is large unless $\phi$
is very small. Furthermore, for small values of $\phi$, $F \simeq 1$.
In fact, for $\sin\phi \ge 10^{-2}$, we find $\gamma \gg 1$ so that the
Landau-Zener probability is vanishingly small: $P_{LZ} < 10^{-2}$.
Our three-flavour analysis of the
neutrino mixing problem in any case precludes the choice of $\phi = 0$.
Therefore, for all practical purposes, we assume $ \phi > 10^{-2}$ and
hence consider the neutrino propagation in matter to be purely adiabatic.
This implies that the
$\vert \nu_e \rangle $ which is produced as $\vert \nu_3 \rangle$
essentially remains in this mass eigenstate until it
reaches the detector. For $\overline{\nu}_e$, which are produced mainly
in the $\vert \overline{\nu}_1 \rangle$ mass eigenstate, there is no
resonance condition to be satisfied (the sign of $A$ changes from
neutrino to antineutrino) and hence the propagation is always adiabatic.

\newpage
\begin{table}[htp]
\begin{tabular}{|r|rrrrrrrr|}
Detector & $\nu_e\,e$ & $\nu_e\,O$ & $\overline{\nu}_e\,e$ &
$\overline{\nu}_e\,O$ & $\overline{\nu}_e\,p$ & $\nu_{\mu,\tau}\,e$ &
{$\overline{\nu}_{\mu,\tau}\,e$} & ~~\\ \hline
 &  \multicolumn{7}{c}{$E_e > 8$ MeV} & \\
1 KTon (no osc) &
    2.3   &  1.0  &   0.8 &    3.8 &  272.0 &    1.3 &    1.0 & \\
1 KTon (max osc) &
    4.2   & 23.8  &   0.9 &    8.1 &  323.2 &    1.0 &    1.0 & \\
SuperK (no osc) &
   72.4   & 30.8  &  25.1 &  123.0 & 8702.9 &   41.1 &   33.4 & \\
SuperK (max osc) &
  134.6   &761.0  &  30.0 &  260.6 &10343.9 &   31.5 &   31.7 & \\ \hline
 &  \multicolumn{7}{c}{$E_e > 30$ MeV} & \\
1 KTon (no osc) & 0.0 & 0.2 & 0.0 & 1.8 & 76.6 & 0.2 & 0.2 & \\
1 KTon  (max osc) & 0.7 & 18.4 & 0.1 &  5.4 & 138.6 & 0.1 & 0.1 & \\
SuperK (no osc) & 
    1.5  &   5.9  &   1.0 &   56.2 & 2450.0   &  6.4  &   4.9 & \\
SuperK (max osc) & 
   22.2  & 587.6  &   2.2 &  172.8 & 4436.5   &  3.5  &   4.1 & \\
\end{tabular}
\caption{Event rates with and without oscillation for a supernova
explosion at a distance of 10 KPc. The high energy events, which are given
separately, show the enormous enhancement in the $p$ and $O$
channels, with oscillation. Results are shown for a 1 kTon water
Cerenkov as well as for the 32 kTon (fiducial volume) SuperKamiokande
detectors.}
\end{table}

\newpage
\begin{figure}[htp]
\centering
\vskip 8truecm

{\includegraphics{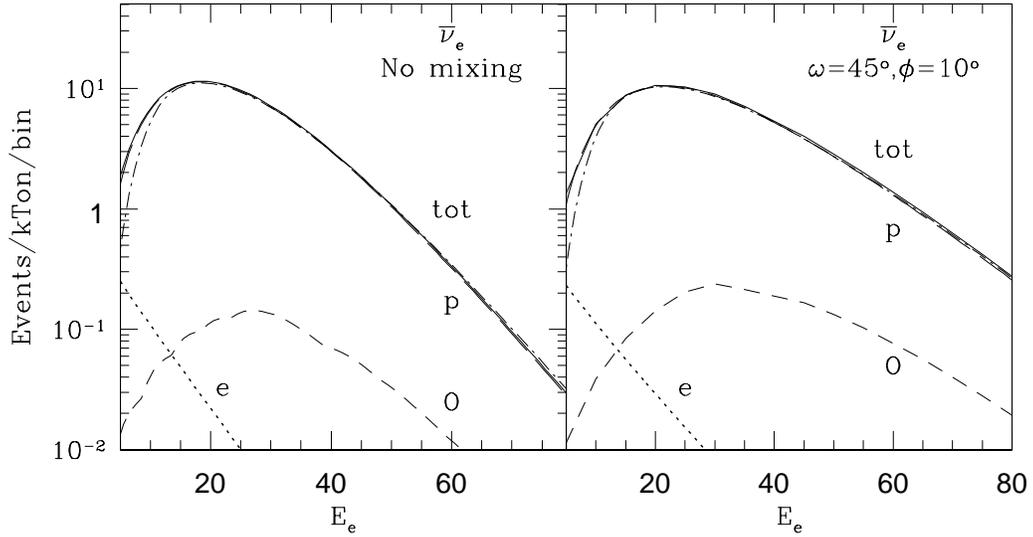}}

\caption{The number of events in bins of electron energy of 1 MeV each,
due to $\overline{\nu}_e$ interactions, are shown as a function of the
electron energy, with and without mixing. The long-dashed, dashed, and
dotted lines correspond to interactions with $p$, $O$ and $e$
respectively, in the detector. The dot-dashed line indicates the effect
of inclusion of detector efficiency and resolution on the interaction
with $p$. See text for more details.  The solid line denotes the total
contribution to the event rate from $\overline{\nu}_e$.}
\end{figure}

\begin{figure}[hbp]
\centering
\vskip 8truecm

{\includegraphics{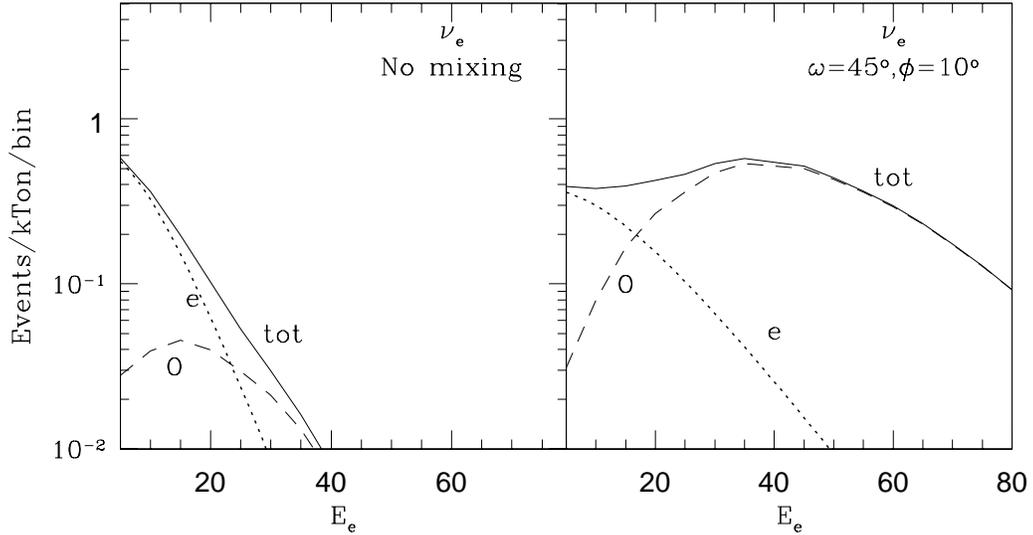}}
\caption{The number of events in bins of 1 MeV each, due to $\nu_e$
interactions, are shown as a function of the electron energy, with and
without mixing. The dashed and dotted lines correspond to interactions
with $O$ and $e$ respectively, in the detector. The solid line denotes
the total contribution to the event rate from $\nu_e$.}
\end{figure}

\begin{figure}[p]
\centering
~ 
\vskip 8truecm

{\includegraphics{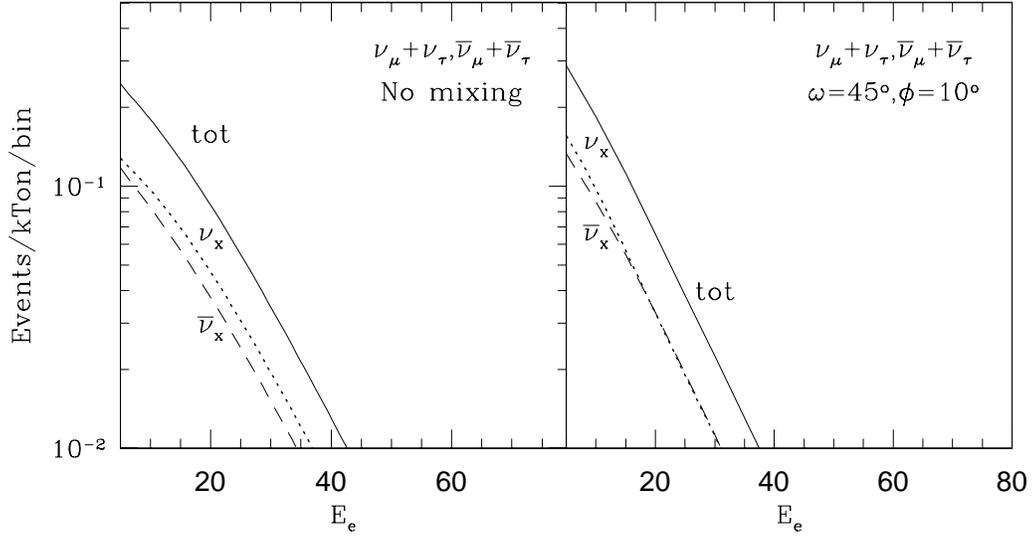}}
\caption{The number of events in bins of 1 MeV each, with and without
mixing, due to $\nu_{\mu,\tau}\,e$ and $\overline{\nu}_{\mu,\tau}\,e$
elastic scattering in the detector, are shown as a function of the
electron energy, as dotted and dashed lines respectively. The solid
line denotes the total contribution to the event rate from all these
channels, that is from $\nu_\mu$, $\nu_\tau$, $\overline{\nu}_\mu$, and
$\overline{\nu}_\tau$.}

\end{figure}

\begin{figure}[p]
\centering
\vskip 9truecm

{\includegraphics{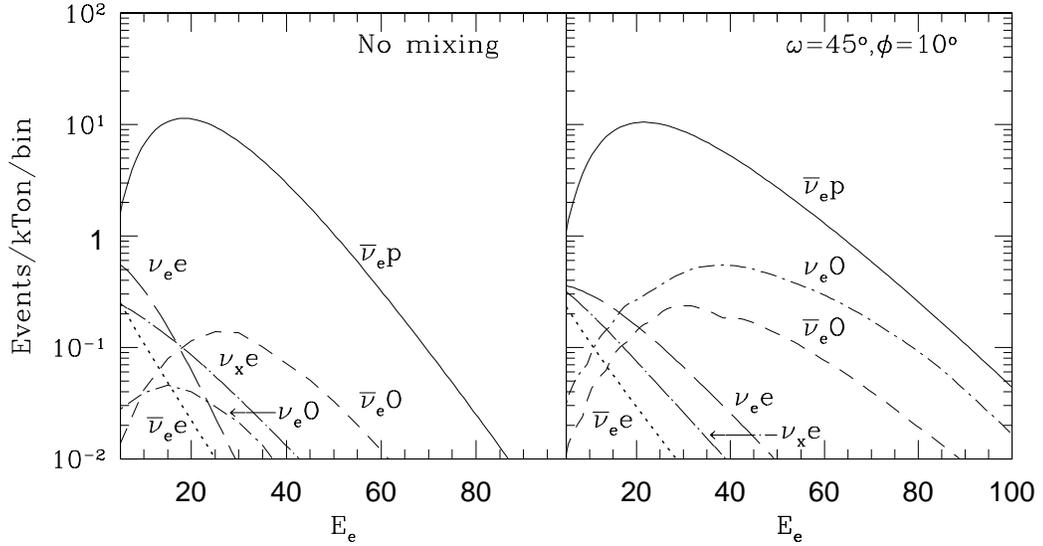}}
\caption{A comparison of the number of events in bins of 1 MeV each,
due to various processes, is shown as a function of the electron
energy, with and without mixing.  The line types indicate events from
the processes $\overline{\nu}_e\,p$ (solid), $\overline{\nu}_e\,e$
(dotted), $\overline{\nu}_e\,O$ (dashed), ${\nu}_e\,e$ (long-dashed),
${\nu}_e\,O$ (dot-dashed), and ${\nu}_x\,e$ (dot-long dashed) processes
respectively. The subscript $x$ denotes the NC contribution from
$\nu_\mu$, $\nu_\tau$, and their antiparticles.}
\end{figure}

\newpage
\begin{figure}[p]
\centering
~
\vskip 18truecm

{\includegraphics{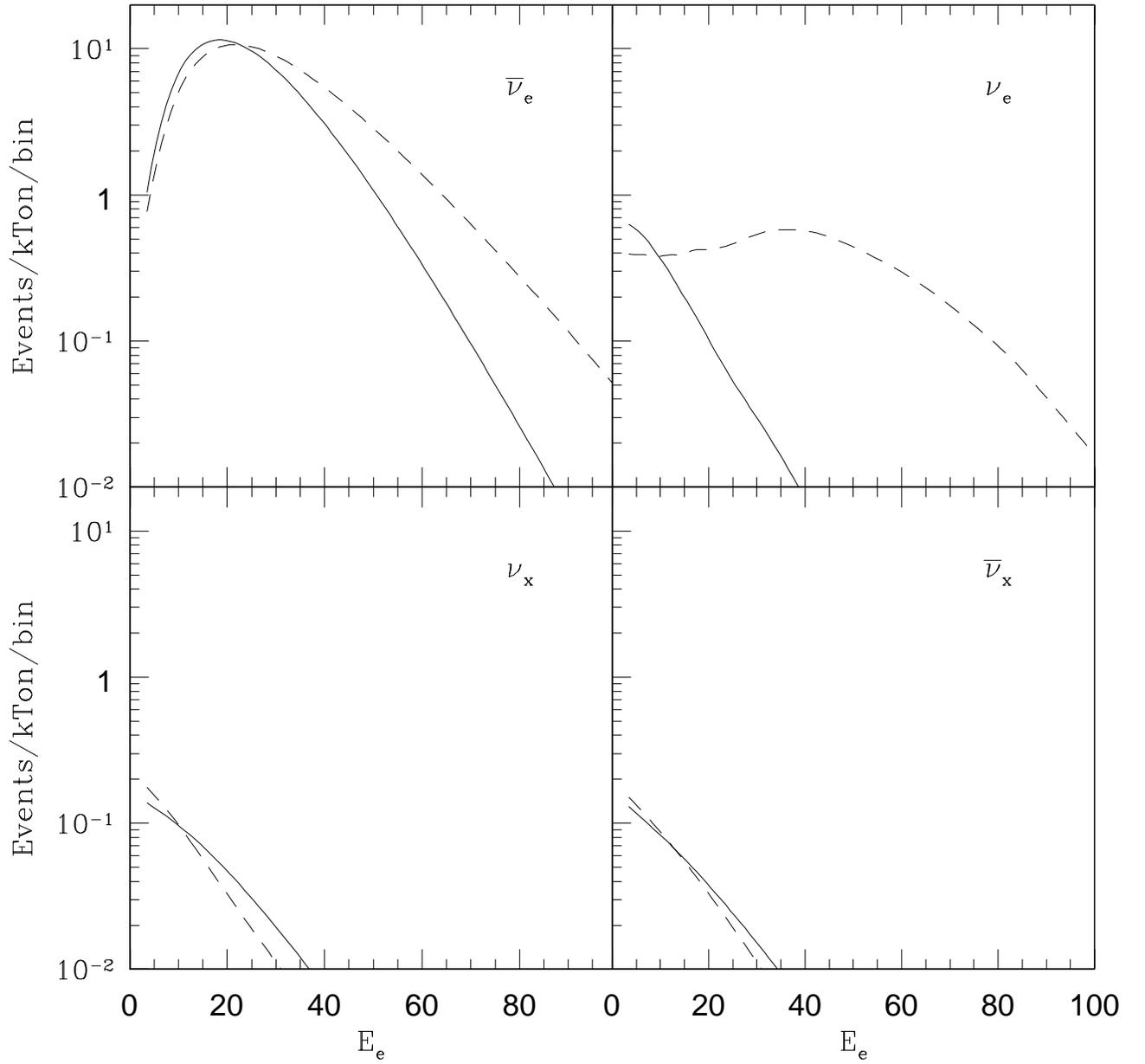}}
\caption{The total number of events (summed over all processes) in bins
of 1 MeV each, due to $\overline{\nu}_e$, $\nu_e$, $\nu_{\mu,\tau}$ and
$\overline{\nu}_{\mu,\tau}$ interactions,  are shown as a function of
the electron energy. The solid and dashed lines denote the event rates
without and with (maximal effect due to) mixing.}

\end{figure}

\newpage
\begin{figure}[p]
\centering
~
\vskip 8truecm

{\includegraphics{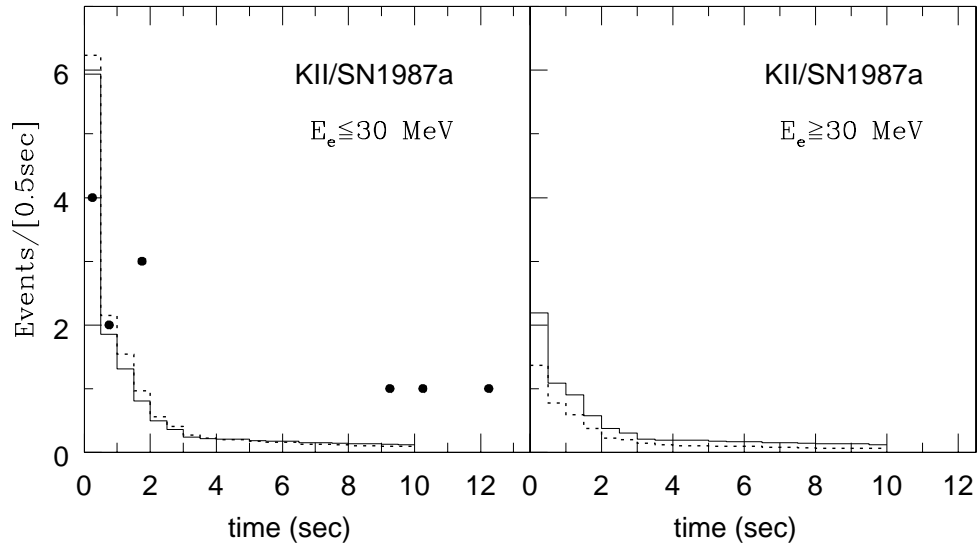}}
\caption{The time-dependent neutrino spectrum (due to
$\overline{\nu}_e\,p$ scattering) in bins of 0.5 sec is shown as a
function of the time of detection, in comparison with the events
observed at the Kamiokande II detector from the supernova SN1987a.  The
dashed and solid lines correspond to the number of events without and
with (maximal effect due to) mixing. The low and high energy components
of the signal are separately shown.}

\end{figure}

\end{document}